\documentclass[aps,twocolumn,showpacs,preprintnumbers,floatfix,amsmath,amssymb]{revtex4-1}

\usepackage{hyperref}

\usepackage{tikzexternal}
\tikzset{external/system call={pdflatex \tikzexternalcheckshellescape -halt-on-error -interaction=batchmode -jobname "\image" "\texsource" && pdftops -eps "\image".pdf}}

\usepackage{color}

\newcommand{\R}{\mathbb{R}}
\renewcommand{\>}{\rangle}
\newcommand{\<}{\langle}

\newcommand{\x}{{\mathbf x}}
\renewcommand{\k}{{\mathbf k}}
\newcommand{\Tr}{\mathop{\text{Tr}}}
\newcommand{\D}{\mbox{$\not\!\!D$}} 

\begin{document}

\title{Cancellation of quantum anomalies and bosonization of \\
three-dimensional time-reversal symmetric topological insulators}

\author{Heinrich-Gregor Zirnstein}
\affiliation{Institut f\"ur Theoretische Physik, Universit\"at Leipzig, D-04103 Leipzig, Germany}
\author{Bernd Rosenow}
\affiliation{Institut f\"ur Theoretische Physik, Universit\"at Leipzig, D-04103 Leipzig, Germany}

\date{July 6th, 2013}


\begin{abstract}
The strong time-reversal symmetric (TRS) topological insulator (TI) in three space dimensions features gapless surface states in the form of massless Dirac fermions. We study these surface states with the method of bosonization, and find that the resulting bosonic theory has a topological contribution due to the parity anomaly of the surface Dirac fermions. We argue that the presence of a quantum anomaly is, in fact, the main reason for the existence of a surface state, by the principle that anomalies of a surface and bulk must cancel. Inspecting other classes of topological insulators, we argue that this principle holds in general. Moving beyond purely topological considerations, we incorporate the dynamics of the surface electron states into the bosonic theory. Additionally, we discuss the thermodynamics of the bosonic theory and propose a representation of the surface Dirac fermions in terms of the bosonic fields.
\end{abstract}

\maketitle

\section{Introduction}
Topological insulators \citep{Hasan:2010ku,Qi:2011hb,PSSR:PSSR201206416} are a recently discovered class of materials whose most salient feature is that they are insulating in the bulk but feature gapless, topologically protected surface states.
Here, we will focus on the so-called ``time-reversal symmetric (TRS) topological insulator (TI) in three space dimensions'' \citep{Fu:2007io,Moore:2007gq,Zhang:2009ks,Xia:2009fn,Brune:2011hi}, which is protected against time reversal invariant perturbations.
In the regime of low energies, the surface states of the TRS topological insulator in three space dimensions are described by $(2+1)$-dimensional massless Dirac fermions \citep{Fu:2007ei,Zhang:2009ks}, which obey the massless Dirac equation $\gamma^\mu\partial_\mu \psi = 0, \mu=0,1,2$. Within this fermionic approach,  the effect of electron interactions on the magnetic susceptibility and on 
magnetic symmetry breaking has been studied  \citep{2009PhRvL.103z6801F,2011PhRvB..83t5124J,2012PhRvB..85l1105B,2012PhRvB..86s5116B}.

In this article, we  discuss a description of TI surface states in terms of a bosonic field theory, which contains both a universal topological term and 
a non-universal term describing the electron-hole continuum of massless Dirac electrons. The idea of describing fermionic systems in terms of composite bosonic excitations, such as the electron density, is known as \emph{bosonization} and has been successfully applied to $(1+1)$-dimensional systems \citep{vonDelft:1998ff,1975PhRvD..11.3026M,1975PhRvD..11.2088C}; here we argue that it is useful in $(2+1)$-dimensional systems as well.

More specifically, we will consider the response of topological insulators to external gauge fields\citep{Qi:2008eu, 2013EL....10147007F} and apply functional bosonization \citep{1994NuPhB.421..373B,Fradkin:1994gk,1999PhRvD..59j5012B,2013PhRvB..87h5132C,1995PhRvB..5210877K} to construct a topological field theory \citep{Cho:2011bd,2013PhRvB..87h5132C} that connects bulk and surface states in a satisfying manner. 
We emphasize the important role of the \emph{parity anomaly} \citep{Redlich:1984ck,Redlich:1984hu,Mulligan:2013te} and of quantum anomalies in general \citep{2012PhRvB..85d5104R,Hughes:2012tc,Furusaki:2012tg,Ringel:2012uo,2007PhRvL..99k6601R,Stone:2012wg} for understanding the existence and topological nature of surface states, and show how the dynamics of the surface states can be incorporated into the field theory. By construction, the resulting bosonic field theory correctly  describes the electromagnetic response in the long wavelength limit, but reproduces the fermionic thermodynamics only qualitatively. We explain that functional bosonization cannot be expected to quantitatively reproduce thermodynamics quantities from the bosonic effective action alone, and that the difference between fermionic and bosonic thermodynamics is accounted  for by normalization factors of auxiliary functional integrals.
Moreover, we discuss the current status of the \emph{refermionization} \citep{vonDelft:1998ff,1979PhR....49..261L,Marino:1991ti} program in $2+1$ dimensions, which aims at studying the underlying fermion degrees of freedom in terms of the bosonic theory and whose main attraction is that it could simplify the study of strong electron interactions.

The structure of the paper is as follows: First, in Section~\ref{section-quantum-anomalies}, we discuss the quantum Hall effect 
as an example for demonstrating how the existence of surface states follows from the principle that quantum anomalies of both surface and bulk must cancel, in this case the chiral anomaly. Then, we apply this idea to the TRS topological insulator in three space dimensions and argue that here the parity anomaly is responsible for the existence of surface states. 
In Section~\ref{section-topological-field-theory}, we review the method of functional bosonization and apply it to the TRS topological insulator to obtain a topological field theory that connects bulk and surface via the parity anomaly. After that, in Section~\ref{section-surface-dynamics}, we focus on the surface states and incorporate the electron dynamics into the bosonic field theory by considering the non-topological parts of the effective electromagnetic response. In Section~\ref{section-surface-thermodynamics}, we calculate the  partition function of both the bosonic and fermionic surface theories, and discuss the discrepancy in the specific heat between the two approaches. 
Finally, in Section~\ref{section-refermionization}, we review and use the so-called tomographic transform to propose a representation of the electron field in terms of the bosonic theory, and a 
summary in Section~\ref{section-summary} concludes the main text.
Appendix \ref{section-calculation} collects some tools for manipulating the surface Chern-Simons term. In Appendix \ref{section-axial-anomaly}, we return to quantum anomalies and discuss how the quantum spin Hall effect relates to the axial anomaly. Finally, Appendix \ref{section-quantum-regularization} gives a very short general introduction to quantum anomalies with a focus on the issue of regularization.

\section{Cancellation of quantum anomalies}
\label{section-quantum-anomalies}

Our goal is to describe topological insulators by a field theory of bosons that arise as collective excitations of the underlying electron field. This approach has been very successful for studying the integer and fractional quantum Hall effects \citep{Wen:1992cf}.

Such a field theory is necessarily topological in nature and its defining feature must be that it correctly predicts the topologically protected surface states. We will now argue that the main mechanism for the emergence of surface states is the existence of  a \emph{quantum anomaly} of the bulk states together with the requirement that this anomaly is \emph{canceled} by the surface states. In particular, for the TRS topological insulator in three space dimensions, the quantum anomaly in question is the \emph{parity anomaly}; 
for other topological phases, the relevant anomalies are different.

In quantum field theory, the notion of a quantum anomaly \citep{Fujikawa:2012uc} refers to the phenomenon that a symmetry, such as gauge invariance or time-reversal symmetry, may be present in the classical Lagrangian, but is lost once the theory is quantized. To make our discussion self-contained, we give a very short introduction to this phenomenon in Appendix~\ref{section-quantum-regularization}.
In particular, this means that the response of the quantum system to an external field, like the electromagnetic field, does not have the expected symmetry either.
We will now describe the response of some classes of topological insulators and see that the introduction of a boundary breaks a symmetry of the bulk, which should be restored by the quantum anomaly of an emerging surface state.


As a warm-up, let us first consider the prototype  example of a topological phase, namely the two-dimensional quantum Hall system. Given an external electromagnetic gauge field $A_\mu$, the real time effective action  of the bulk is of  Chern-Simons type \citep{Wen:1992cf}
\begin{equation}
    \label{eq-action-qhe}
    S_{\text{eff}}[A_\mu] = -\frac{\sigma_{xy}}2 \int d^2x dt\ \varepsilon^{\mu\nu\lambda} A_\mu \partial_\nu A_\lambda
\end{equation}
and describes a Hall response with  Hall conductivity $\sigma_{xy}$.
For periodic boundary conditions, this action is invariant under a gauge transformation $A_\mu \to A_\mu + \partial_\mu \Lambda$, which implies that the current $j^\mu = \frac{\delta S[A]}{\delta A_\mu}$ is conserved, $\partial_\mu j^\mu=0$ as expected.

However, if we restrict this effective action to a domain with boundary, for instance to the half-space $x_1 > 0$, we see that the corresponding action
\begin{equation}
    S_{\text{eff, bulk}}[A_\mu] = -\frac{\sigma_{xy}}2 \int d^2x dt\ \theta(x_1) \varepsilon^{\mu\nu\lambda} A_\mu \partial_\nu A_\lambda
\end{equation}
gives a current that is no longer conserved
\begin{equation}
    \label{eq-bulk-hall-current}
    \partial_\mu j_{\text{bulk}}^\mu = \partial_\mu \frac{\delta S_{\text{eff}}[A]}{\delta A_\mu}
        = -\frac{\sigma_{xy}}2 \delta(x_1) \varepsilon^{1\nu\lambda} \partial_\nu A_\lambda
.\end{equation}
This is not unexpected: a Hall current flows towards the edge of the sample, but cannot continue in the vacuum, so there is a defect at the edge. Of course, it is a general physical principle that the total current must be conserved and we conclude that there \emph{must exist} an edge state that compensates the loss: \citep{Wen:1992cf,2011CRPhy..12..332B}
\begin{equation}
    \label{eq-edge-hall-current}
    \partial_\mu j_{\text{edge}}^\mu = +\frac{\sigma_{xy}}2 \delta(x_1) \varepsilon^{1\nu\lambda} \partial_\nu A_\lambda
\end{equation}
\begin{equation}
    \partial_\mu (j_{\text{bulk}}^\mu + j_{\text{edge}}^\mu) = 0
.\end{equation}

Note the peculiar fact that the fermionic theory at the edge individually also violates current conservation. This is the \emph{chiral anomaly} \citep{Fujikawa:2012uc} in $1+1$ dimensions, and the compensation presented here is the Callan-Harvey mechanism \citep{1985NuPhB.250..427C}. Of course, the response of the total system does conserve the charge current; it is just the individual parts that are anomalous separately.

In summary, even though the anomalous response of the bulk is canceled in the total response, its indirect physical consequences are nonetheless profound: it predicts the \emph{existence} of an edge state. Such an edge state usually carries additional longitudinal responses which can be measured experimentally, for example, charge transport in parallel to the boundary of a quantum Hall sample. 
Thus, looking only at the low-energy effective action \eqref{eq-action-qhe} which abstracts away the microscopic details of the electron system and only captures the bulk electromagnetic response, the principle of anomaly cancellation still allows us to show that electronic states must exist at the boundary. This is the main prediction that we expect from a successful topological field theory for topological insulators.

In the case of the quantum Hall effect, a more direct interpretation of the anomaly is also available. Namely, applying an electric field in parallel to the boundary of a quantum Hall sample will generate a Hall current in the bulk which leads to a transport of charge from the bulk region to the boundary region of the sample \footnote{However, note that attributing the charge in the boundary \emph{region} to either the boundary \emph{states} or the insulating bulk \emph{states} is subtle. In equation \eqref{eq-edge-hall-current}, the prefactor of $\sigma_{xy}/2$ means that only half of the expected Hall charge in the boundary region is attributed to the boundary states, the rest is carried by the gapped bulk states. This attribution corresponds to the \emph{consistent} anomaly. In contrast, the \emph{covariant} anomaly is a formulation where all the charge in the boundary region is attributed to the boundary response. See also Ref.~\citep{1994PhRvD..49.1980C}.}. This has a profound consequence when the boundary of the sample is multiply connected, for example in a Corbino geometry. Then, threading a flux quantum through the inner hole of the Corbino disk leads to charge transport from the inner to the outer edge or vice versa.


We now present a similar argument for the existence of surface states of the strong TRS topological insulator in three space dimensions. This time, however, the surface states are governed by the \emph{parity anomaly} \citep{Redlich:1984ck,Redlich:1984hu} and the physical symmetry principle we invoke is not current conservation (which follows from gauge symmetry), but \emph{time-reversal symmetry}.

The bulk electromagnetic response of a TRS topological insulator in three dimensions with periodic boundary conditions has been derived \citep{Qi:2008eu,Essin:2010gy} as
\begin{align}
    \label{eq-parity-bulk}
    S_{\text{eff}}[A_\mu]
    &= \frac{\theta}{2\pi}\frac{e^2}{hc} \int d^3xdt\ \mathbf{E}\cdot\mathbf{B} + \dots
    \\
    &= -\frac{\theta}{8\pi^2} \int d^3xdt\ \varepsilon^{\mu\nu\rho\sigma}\partial_\mu A_\nu \partial_\rho A_\sigma + \dots
\end{align}
where we have absorbed \footnote{In this article, we absorb the electron charge $e$ into the gauge field $A_\mu$ and choose units such that $\hbar=1$ and $c=1$. In particular, $h=2\pi$ which explains the occasional factor of $2\pi$ in various actions. For reference, the Hall current of a quantum Hall system with $\nu$ filled Landau levels is $j^1 = \sigma_{xy}\mathbf E_2 = -\nu/2\pi \cdot \varepsilon^{1\nu\lambda}\partial_\nu A_\lambda$.} the electron charge $e$ into the definition of the gauge field $A_\mu$. Displayed is the topological part of the response, the dots indicate that there may be additional, non-topological contributions.
Parameter values $\theta = \pm\pi$ indicate a  topologically non-trivial insulator, $\mathbf E$ and $\mathbf B$ are the electric and the magnetic field respectively. It is not obvious why this action should be time-reversal invariant, since after all, time reversal maps $\mathbf{E}\cdot\mathbf{B} \to -\mathbf{E}\cdot\mathbf{B}$. However, for periodic boundary conditions, the integral $\int d^3xdt\ \varepsilon^{\mu\nu\rho\sigma}\partial_\mu A_\nu \partial_\rho A_\sigma$ is actually quantized to an integer multiple of $8\pi^2$, this is a general property of gauge connections related to the so-called second Chern class \citep{Nakahara:2006vx}. Thus, time-reversal does change the action by $\Delta S = 2\theta n = 2\pi n$, but only the Feynman path amplitude $e^{iS}$ has physical meaning in quantum theory, and this one remains invariant.

Once again, let us now restrict the action to a domain $\Omega$ with boundary $\partial\Omega$. The bulk action can be written as a boundary integral 
\begin{align}
    S_{\text{bulk}}[A_\mu]
    &= -\frac{\theta}{8\pi^2} \int_\Omega d^3xdt\ \varepsilon^{\mu\nu\rho\sigma}\partial_\mu A_\nu \partial_\rho A_\sigma
    \\ \label{eq-chern-boundary}
    &= -\frac{\theta}{8\pi^2} \int_{\partial\Omega} d^2xdt\ \varepsilon^{\nu\rho\sigma}A_\nu \partial_\rho A_\sigma
\end{align}
which is no longer quantized or time-reversal invariant. But since the total action of the system must be time-reversal invariant, there must exist a surface state whose response makes the total action $S_{\text{bulk}}+S_{\text{surface}}$ time-reversal invariant, for instance by
\begin{equation}
    \label{eq-parity-surface}
    S_{\text{surface}}[A_\mu] = \frac{\theta}{8\pi^2} \int_{\partial\Omega} d^2xdt\ \varepsilon^{\nu\rho\sigma}A_\nu \partial_\rho A_\sigma + 
   \dots
.\end{equation}
This effective action precisely matches the parity anomaly \citep{Redlich:1984ck,Redlich:1984hu} of the $(2+1)$-dimensional massless Dirac fermion, which is  the surface state that we would obtain by solving a fermionic (model) Hamiltonian directly. Again, this is the topological part of the action, the dots indicate that other, time-reversal invariant contributions are possible, for instance those that describe the dynamics of the surface state. We will discuss them in Section~\ref{section-surface-dynamics}.

This topological contribution of the surface looks like a quantum Hall effect on the boundary. However, we stress  that it cancels out in the total electromagnetic response and cannot be observed experimentally. This was recently also shown by an explicit computation of the fermion Green's function in the presence of a boundary \citep{Mulligan:2013te}. In other words, the bulk action $(\ref{eq-parity-bulk})$ correctly describes a TRS topological insulator with \emph{gapless} surface states, in combination with the anomaly cancellation mechanism presented here. There is no need to artificially break time-reversal symmetry in order to describe TRS topological insulators.

In summary, the most important if indirect physical consequence of the parity anomaly is the existence of electronic surface states, which have unusual longitudinal response that can be measured experimentally. However, unlike for the quantum Hall system, we are not aware of a more direct interpretation of the parity anomaly, because the charge current obtained from the action \eqref{eq-chern-boundary} is localized on the boundary and does not flow in the bulk region. For instance, for a topological insulator in the region $x_1>0$, the bulk current is calculated as
\begin{equation}
    j_{\text{bulk}}^\mu = -\frac{\theta}{4\pi^2}
        \delta(x_1) \epsilon^{1\mu\nu\lambda} \partial_\nu A_\lambda
\end{equation}
which is canceled completely by the boundary current $j^\mu_{\text{surface}}$.
Thus, unlike in the quantum Hall case, we do not believe that an experiment involving only electromagnetism can find a direct signature of the parity anomaly of the gapless topological insulator. Moreover, the parity anomaly already captures the topological essence of the situation, namely the competition between time-reversal symmetry and charge conservation, so we expect that trying to find a direct signature of the parity anomaly by measuring a different quantity (like spin) remains a challenge for the future. 


In general, we believe that this mechanism of anomaly cancellation applies to all topological insulators with any symmetry and in any dimension. The gapless states on the boundary should always feature a quantum anomaly that cancels with the bulk; this is why they exist in the first place. Recent theoretical evidence \citep{2012PhRvB..85d5104R,Hughes:2012tc,Furusaki:2012tg,Mulligan:2013te,Ringel:2012uo,2007PhRvL..99k6601R,Stone:2012wg} has established a general connection between topological insulators and quantum anomalies. Importantly, all symmetry classes \citep{PSSR:PSSR201206416} of topological insulators are related to anomaly polynomials \citep{2012PhRvB..85d5104R}. Our conjecture about the 
connection between boundary states and quantum anomalies is supported by  the following collection of examples:

i) Integer quantum Hall effect in $2$ space dimensions, symmetry class A. 
Transfer of charge between bulk and edge due to the \emph{chiral anomaly} of the boundary states.

ii) TRS topological insulator in $3$ space dimensions, symmetry class AII. Cancellation of the \emph{parity anomaly} on the boundary as discussed above.

iii) Quantum spin Hall insulator in $2$ space dimensions, symmetry class AII. Combined spin or charge flow between edge and bulk  due to the \emph{axial anomaly} \citep{Jackiw:2008fg}. We delegate the discussion of this instructive example to Appendix \ref{section-axial-anomaly} as we want to focus on three space dimensions in the main text.

iv) Particle-hole symmetric topological superconductor in $2$ space dimensions, symmetry class D. The boundary states are described by Majorana fermions, but these do not carry charge, so they cannot give rise to a quantum anomaly involving the electromagnetic field. The boundary modes do, however, conduct heat and it has recently been shown \citep{Hughes:2012tc} that the (thermal) energy and momentum current has a quantum anomaly and flows from the bulk into the edge. As energy and momentum are the conserved quantities corresponding to spacetime translations, they can be probed by distorting spacetime, i.e.~by introducing a gravitational field \footnote{Gravitational fields may seem somewhat unusual in a condensed matter setting, but they are useful for calculating response functions of the \emph{energy} and \emph{momentum currents}. One could picture a gravitational field as a deformation of the crystal lattice, but we rather take the point of view that gravitational fields are best seen as a convenient means to obtain the energy-momentum current from Noether's theorem by a variation of the action with respect to the ``spacetime distortion field'' $g_{\mu\nu}$ as $T_{\mu\nu} = \frac{\delta S[g]}{\delta g_{\mu\nu}}$, similar to how the electric current can be obtained by a variation with respect to the electromagnetic field as $j^\mu = \frac{\delta S[A]}{\delta A_\mu}$.} Hence, it is a \emph{gravitational anomaly} \citep{Stone:2012wg, Hughes:2012tc, 1984NuPhB.234..269A} that explains the emergence of Majorana fermions at the boundary.

We believe that this list of examples is sufficiently diverse 
to support our conjecture about the connection between quantum anomalies and topological surface states, as it includes the basic known quantum anomalies, features different cancellation mechanisms and mentions both primary insulators and secondary descendants obtained by dimensional reduction \citep{Qi:2008eu}.

\section{Topological Field Theory}
\label{section-topological-field-theory}

Having understood the parity anomaly as the topological mechanism that requires the existence of surface states, we can now derive topological field theories for surface and bulk. This can be done with a general method called \emph{functional bosonization} \citep{1994NuPhB.421..373B,1999PhRvD..59j5012B,Fradkin:1994gk,2013PhRvB..87h5132C,1995PhRvB..5210877K}, which we will now review
to make our discussion self-contained. 

The main idea of functional bosonization is to start with some effective action $S_{\text{eff}}[A_\mu]$ for the external vector potential that was obtained by integrating out the fermion degrees of freedom. Then, we subject this action to a so-called \emph{duality transform} \citep{1994NuPhB.421..373B}, which yields bosonic fields that generate the same effective action. These fields can then be interpreted as collective charge excitations of fermions. In a sense, we ``undo'' the process of integrating out the matter fields.

For concreteness, let us imagine the example of a massless $(2+1)$-dimensional Dirac fermion, which represents the surface state. Assume for a moment that we have somehow managed to calculate its partition function and effective action
\begin{multline}
    \label{eq-partition-dirac}
    Z[A_\mu] = \int D\bar \psi D\psi \exp\left(i\int d^2xdt\ \bar \psi i\gamma^\mu(\partial_\mu - iA_\mu)\psi \right)
    \\ = e^{iS_{\text{eff}}[A_\mu]}
\end{multline}
when coupled to an external gauge field $A_\mu$. We know that this functional is gauge invariant, so we could as well average over all configurations that are connected by a gauge transformation
\begin{equation}
    Z[A_\mu] = \int D\Lambda\ Z[A_\mu + \partial_\mu \Lambda]
\end{equation}
up to a normalization factor that we ignore for now.
We can write this as an integration over a field $a_\mu$ that is constrained to be a pure gauge $a_\mu = \partial_\mu \Lambda$. By the Poincar\'e lemma, this constraint is equivalent to the vanishing of the curl $\varepsilon^{\mu\nu\lambda}\partial_\nu a_\lambda$, which we can impose with a delta function
\begin{equation}
    Z[A_\mu] = \int Da_\mu\ \delta(\varepsilon^{\mu\nu\lambda} \partial_\nu a_\lambda) Z[A_\mu + a_\mu]
.\end{equation}
Shifting $a_\mu \to a_\mu - A_\mu$ and enforcing the delta function constraint in terms of a Lagrange multiplier $W_\mu$, we obtain
\begin{align}
   \label{eq-bosonization}
   Z[A_\mu] &= \int Da_\mu DW_\mu\ \exp\Big(iS_{\text{eff}}[a_\mu]
   \\ &\quad \nonumber
   - \frac{i}{2\pi}\int d^2xdt\ \varepsilon^{\mu\nu\lambda}W_\mu \partial_\nu(a_\lambda-A_\lambda)\Big)
   \\
   & =: \int Da_\mu DW_\mu\ \exp\Big(iS_{\text{boson}}[a_\mu,W_\mu,A_\mu]\Big)
.\end{align}
The expression in the exponential is the desired action for the bosonic fields $a_\mu$ and $W_\mu$. The whole procedure is essentially just a fancy way to rewrite our starting point, the effective fermionic action, in terms of the charge current
\begin{equation}
    \label{eq-current}
    j^\mu = \frac{\delta S_{\text{boson}}[a,W,A]}{\delta A_\mu} \equiv \frac{1}{2\pi} \varepsilon^{\mu\nu\lambda} \partial_\nu W_\lambda
,\end{equation}
which has been expressed as the curl of a field $W_\mu$ because it needs to fulfill $\partial_\mu j^\mu=0$.

Applying this method to the analysis of surface states, i.e.~to the effective action $(\ref{eq-parity-surface})$ corresponding to the parity anomaly, we obtain a bosonized theory 
\begin{multline}
\label{eq-top-surface}
      S_{\text{surface}}[a_\mu,W_\mu,A_\mu] = \frac1{2\pi} \int d^2xdt\
\Big[\frac{\theta}{4\pi} \varepsilon^{\mu\nu\rho}a_\mu \partial_\nu a_\rho \\ - \varepsilon^{\mu\nu\rho}W_\mu \partial_\nu a_{\rho} + \varepsilon^{\mu\nu\rho}W_\mu \partial_\nu A_{\rho}\Big]  
.\end{multline}
This is the topological field theory of the surface state; it is equivalent to the celebrated Chern-Simons theory \citep{Wen:1992cf}. It is topological as the corresponding equations of motion for the bosonic fields are devoid of dynamics. Of course, the actual massless Dirac fermions on the surface do have dynamics, i.e.~their effective action contains terms aside from the parity anomaly. We will discuss these shortly, but for now let us focus on the topological aspects following from the parity anomaly.

In the same way, a topological field theory for the bulk can be obtained \citep{2013PhRvB..87h5132C}. In three space dimensions, the derivation just presented has to be changed slightly: the Lagrange multiplier needs to incorporate more constraints to make the field $a_\mu$ a pure gauge, and so will become a tensor field $b_{\mu\nu}$ instead of a vector field $W_\mu$. Starting with the effective bulk action $(\ref{eq-parity-bulk})$, one obtains the topological field theory
 \citep{2013PhRvB..87h5132C}
\begin{multline}
        \label{eq-top-bulk}
    S_{\text{bulk}}[a_\mu,b_{\mu\nu},A_\mu] = \frac1{2\pi} \int_{\Omega} d^3xdt\
\Big[-\frac{\theta}{4\pi} \varepsilon^{\mu\nu\rho\sigma}\partial_\mu a_\nu \partial_\rho a_\sigma +
\\ - \varepsilon^{\mu\nu\rho\sigma} b_{\mu\nu} \partial_\rho a_\sigma + \varepsilon^{\mu\nu\rho\sigma} b_{\mu\nu} \partial_\rho A_\sigma
\Big]
\end{multline}
for the bulk.

We can now connect the topological field theories for bulk and surface. Cancellation of the parity anomaly dictates that the topological bulk and surface actions should cancel, so we have to be able to express the former as an integral over the boundary. This is indeed possible: First, the term $\varepsilon^{\mu\nu\rho\sigma}\partial_\mu a_\nu\partial_\rho a_\sigma$ can be moved to the boundary similar to Equation $(\ref{eq-chern-boundary})$. Second, we can utilize the Euler-Lagrange equations from variation of the field $a_\mu$ to obtain the condition $\varepsilon^{\mu\nu\lambda}\partial_\mu b_{\nu\lambda}=0$. This means that the Lagrange multiplier is forced to be a pure gauge $b_{\mu\nu} = -\frac12 (\partial_\mu W_\nu - \partial_\nu W_\mu)$. Using  this relation in Eq.~(\ref{eq-top-bulk}), we can move the remaining two terms to the boundary as well and we obtain precisely the negative of the surface action.

Compared to previous proposals for field theories of topological insulators \citep{Cho:2011bd, 2013PhRvB..87h5132C},  our approach 
is guided by the principle of anomaly cancellation between bulk and surface states. In contrast to \citep{Cho:2011bd}, we arrive at a 
surface action which is not invariant under time reversal on its own, but together with the  bulk action is time reversal invariant, and elucidates the topological 
meaning of the parity anomaly.

\section{Surface dynamics}
\label{section-surface-dynamics}
While the cancellation of quantum anomalies captures the topological content of the field theories for surface and bulk, it is clear that topology alone cannot give any information about the dynamics of the surface states. For instance, the edge action for the quantum Hall effect may contain any additional term that is compatible with gauge invariance \citep{2011CRPhy..12..332B}. Likewise, the TRS topological insulator in three space dimensions may contain terms compatible with both time-reversal invariance and gauge invariance. Unlike previously suggested \citep{Cho:2011bd}, these terms are not restricted to represent a potential energy. In fact, we will obtain kinetic terms that contain both space and time derivatives, even in a non-local fashion. 

To obtain the dynamic parts of the surface theory, we have to go back to the microscopic theory and use the fact that the electronic states on the surface are described specifically by the massless $(2+1)$-dimensional Dirac fermion. So far, their effective action could not be evaluated exactly, but it has been computed to quadratic order in the field $a_\mu$ \citep{Barci:1996bw, MTsvelik:2005wh,2010PhRvB..82g5133W}. Using this result in the functional bosonization formula \eqref{eq-bosonization}, we obtain the full surface action
\begin{multline}
    \label{eq-surface-full}
    S_{\text{surface}}[a_\mu,W_\mu,A_\mu]
=
\int d^2xdt \left[ \frac{\theta}{8\pi^2}\varepsilon^{\mu\nu\lambda}a_\mu \partial_\nu a_\lambda
\right.\\\left.
+ \frac{1}{16} a^T_\mu(-\square)^{1/2}a^T_\mu + \mathcal{O}(a^4)
\right.\\\left.
- \frac1{2\pi}\varepsilon^{\mu\nu\lambda}(a_\mu-A_\mu)\partial_\nu W_\lambda
\right]
,\end{multline}
\def\x{\mathbf{x}}%
where $a^T_\mu$ denotes the transversal components of the vector field $a_\mu$.
The first term is the by now familiar parity anomaly, while the second one describes the electron dynamics. The dynamic term looks quite unusual, though: It is not a purely potential energy, for it involves the temporal derivatives of the $a_\mu$ field. It is not the Maxwell term either, because it features the square root of the d'Alembert operator $(-\square)^{1/2}$, a non-local pseudo-differential operator. Fortunately, the retarded Green's function $G^R$ of this operator is well defined in $(2+1)$-dimensional spacetime \citep{Edelstein:1996um}
\begin{equation}
    \label{eq-greens}
    G^R(x,x') = \frac{1}{4\pi|\x-\x'|}\delta(t-t'-|\x-\x'|)
.\end{equation}
This is just the $3+1$ dimensional wave propagator restricted to the plane $\x_3=0$. Note that only values on the boundary of the light cone $t-t'=|\x-\x'|$ contribute to the propagator; in mathematics, this property is sometimes called Huygen's principle. Compare this to the usual $(2+1)$-dimensional wave propagator, which is logarithmic and also involves points inside the light cone, not just the boundary.

\def\q{\mathbf q}
In momentum space, in particular in Euclidean spacetime, the square root term is not unusual either, it is simply the magnitude of the Euclidean momentum $(-\square)^{1/2} \cong |q|$. This dynamic term corresponds to the density-density correlation function \citep{2010PhRvB..82g5133W}
\begin{equation}
    \< j^0(-q) j^0(q)\> = \frac1{16}\frac{\q^2}{\sqrt{q_0^2 + \q^2}}
\end{equation}
which yields $|q|$ when combined with the projection onto the transversal components $a^T_\mu(q) = (\delta_{\mu\nu} - \frac{q_\mu q_\nu}{|q|^2}) a_\mu(q)$. Here, the spacetime momentum is $q=(q_0,\q)$.

At this point, we believe  it is useful to discuss the effective action for \emph{massive} $(2+1)$-dimensional Dirac fermions as well, since we feel that there has been some confusion in the literature about the implications of  time-reversal symmetry breaking via a Zeeman field on the surface states. For $(2+1)$-dimensional fermions, the coupling to a Zeeman field has the same form as a mass term, and we will denote the magnitude of the Zeeman energy by $m$ in the following. 
In Euclidean spacetime, the quadratic part of the action has been calculated \citep{Barci:1996bw,MTsvelik:2005wh} in terms of Euclidean momentum $q$ as
\begin{multline}
    \label{eq-surface-massive}
    S_{\text{eff}}[A_\mu] =
    \\ \sum_{q} A_\mu(-q) \Bigg[
    \frac{\theta}{8\pi^2}\varepsilon_{\mu\nu\lambda} q_\lambda
    + \frac{1}{8\pi}\varepsilon_{\mu\nu\lambda} q_\lambda f_1\left(\frac{|q|}{2m}\right)
    + \\ + \left(\delta_{\mu\nu} - \frac{q_\mu q_\nu}{|q|^2}\right)
        \cdot\frac{|q|}{16} f_2\left(\frac{|q|}{2m}\right) \Bigg]A_\nu(q)
    + \mathcal{O}(A^3)
\end{multline}
%
where $m$ is the fermion mass, $f_1(x) = \tan^{-1}(x)/|x|$ and $f_2(x)$ behaves like $x$ for small $x \ll 1$ but approaches $1$ for large $x \gg 1$.
[Equation \eqref{eq-surface-full} corresponds to the special case $m=0$.]
The first term is the familiar parity anomaly, the second term is an additional surface Hall effect, and the final term describes the fermion dynamics.
The mass $m$ of the fermions sets an energy scale. For momenta small compared to the mass $|q| \ll m$, the dynamics resemble a Maxwell term with a coefficient $\frac1{m}$. 
In addition, a surface Hall effect with a Hall conductivity $\sigma_{xy} = \frac12 \frac{m}{|m|}\frac{e^2}h$ is generated,   which corresponds to 
 the zero-energy  Landau level of relativistic fermions. However, for momenta large compared to the mass, $|q| \gg m$, the square root of the wave operator $(-\square)^{1/2}\cong |q|$ dominates the dynamics and the Hall-type term vanishes. Whether a Hall effect can be measured in a static charge transport experiment \citep{Sitte:2012ib,Brune:2011hi} is now a delicate question as the static limit $|q|\to 0$ must be considered in relation to the scales given by the Zeeman magnetic field $m$ and the temperature. In any case, 
we want to emphasize  that the surface Hall effect induced by the mass is independent of the parity anomaly, as the latter has its origin in the need to regularize at large momenta and is not observable, because it cancels with the bulk. A more detailed discussion of the derivation of the effective action \eqref{eq-surface-massive} and the origins of the parity anomaly and the surface Hall effect can be found in Appendix~\ref{section-quantum-regularization}.

It seems somewhat unsatisfactory that we had to go back to the microscopic theory of the surface states and could not make an educated guess to get the non-local term describing the electron dynamics directly from the topological field theory $(\ref{eq-top-surface})$. (In contrast, for the quantum Hall effect, one can add a simple potential energy or use a generalized gauge fixing to get the right edge dynamics.) However, it has been noted \citep{1999PhRvD..59j5012B} that the parity anomaly subsumes the dynamic term, at least in the following sense: \emph{Any} term that is gauge invariant can be absorbed into the Chern-Simons term by means of a non-linear field redefinition. Let us illustrate this for our particular term that describes the dynamics. Define a new field $\tilde a_\mu$ such that
\begin{align}
    \tilde a^T_\mu &= \cos \beta\cdot a^T_\mu + \sin \beta \cdot\varepsilon_{\mu\nu\lambda} \partial_\nu (-\square)^{-1/2} a^{T}_{\lambda}
    \\
    \tilde a^L_\mu &= a^L_\mu
\end{align}
with $\beta$ a complex number such that $\sin(2\beta) = \frac{\pi^2}{2\theta}$. Using the calculations in Appendix~\ref{section-calculation}, we see that the Chern-Simons term for this new field will expand into the parity anomaly plus the dynamic term for the old field
\begin{multline}
    \frac{\theta}{8\pi^2}\tilde a_\mu \varepsilon^{\mu\nu\lambda} \partial_\nu\tilde a_\lambda
    =
    \frac{\theta}{8\pi^2}\varepsilon^{\mu\nu\lambda}a_\mu \partial_\nu a_\lambda + \frac{1}{16} a^T_\mu(-\square)^{1/2}a^T_\mu
.\end{multline}
Even beyond second order in perturbation theory, it is always possible to find a field $\tilde a_\mu$ depending in a non-linear way on $a_\mu$ such that the action becomes the Chern-Simons term for the $\tilde a_\mu$ field \citep{1999PhRvD..59j5012B}. In this sense, the surface action is given by a single Chern-Simons term arising from the parity anomaly, at least modulo a field redefinition. However, we note that such a redefinition, being non-linear in general, gives rise to a non-trivial Jacobian in the path integral.

The bosonic action presented here may also provide a good starting point for studying the effect of electron interactions. After all, the electron density is given by $j^0 = 1/2\pi (\partial_1W_2 - \partial_1W_2)$ and we can model for instance the Coulomb interaction by adding a term
\begin{multline}
    S_{\text{Coulomb}}[W_\mu] = \frac1{4\pi^2}\int dtd^2xd^2y\ (\partial_{x_1}W_2 - \partial_{x_2}W_1)
    \cdot\\
    \cdot\frac1{|\mathbf x - \mathbf y|}(\partial_{y_1}W_2 - \partial_{y_2}W_1)
\end{multline}
to the bosonic action. Integrating out the $a_\mu$ field gives an action that is again quadratic in the $W_\mu$ field and hence exactly solvable. Of course, this assumes that we expand the effective electron action only to quadratic order, i.e.~that we stick to a random phase approximation.

\section{Surface Thermodynamics}
\label{section-surface-thermodynamics}

While the bosonized action $(\ref{eq-surface-full})$ describes the low-energy dynamical and topological properties of the surface electrons correctly, we would like to remark that --- unlike in $1+1$ dimensions --- thermodynamic quantities like the specific heat are not captured by this action. After all, the field $W_\mu$ is just a Lagrange multiplier and the field $a_\mu$ is only a gauge degree of freedom, and hence their partition function is trivial from a thermodynamic point of view. Of course, the total partition function still contains the fermionic thermodynamics, but one needs to carefully keep track of prefactors in the bosonization procedure to see this. In contrast, in  $1+1$ dimensions these prefactors can be interpreted as a  fermionic determinant. As a consequence, 
the fermionic and bosonic action give rise to the same free energy. 

For completeness, let us present the calculation in some detail. First, we consider a massless Dirac fermion. The logarithm of its partition function can be obtained from the functional field integral \citep{Kapusta:2006th} as
\begin{align}
    \nonumber
    \ln Z_{\text{Fermi}} &= \ln \det (\gamma^\mu \partial_\mu)
    \\ &= \frac12 \ln \det (-\partial^2) = \Tr_q \ln (\beta^2|q|^2)
    \label{freenergyfermi.eq}
\end{align}
where $\beta=1/kT$ is the inverse temperature, $q=(q_0,\mathbf q)$ denotes Euclidean momentum and $q_0 = \varepsilon_n \equiv (2n+1)\pi/\beta$ is a fermionic Matsubara frequency. In the last equality in Eq.~$(\ref{freenergyfermi.eq})$, the momentum $|q|^2$ has gained a factor of $\beta^2$ due to the change of the path integral measure under Fourier transformation of the field variable.
A standard calculation \citep{Kapusta:2006th,Altland:2010ww} yields the internal energy
\begin{align}
    U_{\text{Fermi}}
    &= \nonumber
    -\frac{\partial}{\partial\beta} \ln Z
    = 2V \int \frac{d^2\q}{(2\pi)^2} |\q|\frac{1}{e^{\beta|q|}-1}
    \\ &= VT^3\frac{\zeta(3)}{\pi}\frac32
    \label{eq-fermion-energy}
\end{align}
and hence the specific heat
\begin{equation}
    c_{V,\text{Fermi}}
    = \frac1V \left(\frac{\partial U}{\partial T}\right)_V
    = 3T^2\frac{\zeta(3)}{\pi}\frac32
.\end{equation}

Now, let us compute the partition function for the bosonic action $(\ref{eq-surface-full})$. To this end, we have to express it in terms of the Euclidean momentum $q$, similar to Eq.~$(\ref{eq-surface-massive})$. Since this will prove useful again later, we have moved the corresponding formula $(\ref{eq-surface-Euclidean})$ to the Appendix \ref{section-calculation}. Ultimately, the result is that using two matrices $P^+_{\mu\nu},P^-_{\mu\nu}$ that project on the transversal degrees of freedom, we can conveniently express the action as
\begin{equation}
    S[a_\mu,W_\mu] = \sum_q (a_\mu(-q), W_\mu(-q)) D_{\mu\nu}(q)\begin{pmatrix} a_\nu(q) \\ W_\mu(q) \end{pmatrix}
\end{equation}
where
\begin{multline}
    D_{\mu\nu}(q) =
    \\
    |q|\begin{pmatrix}\frac1{16}((1+i\alpha)P^+ + (1-i\alpha)P^-) & -\frac{1}{4\pi} (iP^+ - iP^-) \\ -\frac{1}{4\pi} (iP^+ - iP^-) & 0 \end{pmatrix}
\end{multline}
is a quadratic kernel,  and $\alpha=\frac{2\theta}{\pi^2}$ is a constant related to the prefactor of the Chern-Simons term. Now, we have to project onto the physical transversal degrees of 
 freedom, as the matrix would be singular otherwise. Choosing a convenient basis, we can write the projections as $2\times 2$ matrix blocks to obtain
\begin{align}
    \det \widetilde D_{\mu\nu}(q) &= \det |q| \begin{pmatrix}
        \frac1{16} (1+i\alpha) & 0 & -\frac{i}{4\pi} \\
        0 & \frac1{16} (1-i\alpha) & 0 & \frac{i}{4\pi} \\
        -\frac{i}{4\pi} & 0 & 0 & 0 \\
        0 & \frac{i}{4\pi} & 0 & 0
    \end{pmatrix} \nonumber
    \\ &= \frac1{(4\pi)^4} |q|^4
.\end{align}
We note that the precise numerical value of the $4\times 4$ determinant does not matter as long as it is independent of temperature,  since it contributes only additively to the free energy.  The important result is that   we have four bosonic degrees of freedom with a kernel proportional to $|q|$. From this follows  the partition function
\begin{align}
    \ln Z_{\text{Bose}}
    &=\nonumber \frac12 \Tr_q \ln (\beta^4|q|^4) + \text{const.}
    \\ &= \Tr_q \ln (\beta^2 |q|^2) + \text{const.}
\end{align}
which corresponds to two copies of the canonical Bose gas. Once again, the internal energy is calculated by standard methods to be
\begin{align}
    \label{eq-boson-energy}
    U_{\text{Bose}}
    &=\nonumber
    -\frac{\partial}{\partial\beta} \ln Z
    = 2V \int \frac{d^2\q}{(2\pi)^2} |\q|\frac{1}{e^{\beta|q|}+1}
    \\ &= VT^3\frac{\zeta(3)}{\pi}2 \ \ , 
\end{align}
and this result is larger by a factor of $4/3$ than  the internal energy calculated for the fermions $(\ref{eq-fermion-energy})$.

It may be useful to compare the above calculation to that for the Maxwell-Chern-Simons Theory \citep{1994PhRvD..50.5314B}, whose action is similar to the action $(\ref{eq-surface-full})$ except that our non-local square root $(-\square)^{1/2}$ is replaced by a standard Maxwell term $1/2 a^T_\mu(-\square)a_\mu^{T}$. The main difference is that the Maxwell term is governed by momentum squared $q^2$, while the Chern-Simons term is proportional to the absolute value $|q|$, so that the form of the determinant
is  $\det(q^2+K|q|) = \det(|q|)\det(|q|+K)$, where  the second factor  can be interpreted as arising from   a massive boson. In our case, however, both the dynamics and the Chern-Simons term are governed by $|q|$, so that the surface bosons are related to \emph{massless} canonical bosons.

However, the apparent discrepancy between the fermionic and bosonic specific heat is resolved when taking normalization factors of auxiliary functional 
integrals into account. When dealing with the longitudinal degrees of freedom, it is important to keep track of Fadeev-Popov ghosts, i.e.~functional determinants arising from gauge fixing \citep{Kapusta:2006th}. But since the field $a_\mu$ is only a gauge degree of freedom and $W_\mu$ a Lagrange multiplier, the ghost contributions will conspire to cancel precisely the partition function calculated above. To see this in detail, we have to go back to the functional bosonization procedure $(\ref{eq-bosonization})$ and keep track of prefactors. In particular, averaging over the gauge freedom becomes
\begin{align}
    Z[A_\mu]
&=\nonumber
N_\Lambda \int D\Lambda\ Z[A_\mu + \partial_\mu \Lambda]
\\ &=\nonumber
N_\Lambda \int D\Lambda Da_\mu\ \delta(a_\mu-\partial_\mu\Lambda) Z[A_\mu + a_\mu]
\\ &=\nonumber
N_\Lambda \int D\Lambda Da_\mu\ \delta\left(\Lambda-\frac{a_0}{\partial_0}\right)\det(\partial_0)\cdot
\\ &\qquad\nonumber
\cdot\delta(\partial_0a_1-\partial_1a_0)\delta(\partial_2a_0-\partial_0a_2) Z[A_\mu + a_\mu]
\\ &=
N_\Lambda \int Da_\mu\ \det(\partial_0)\cdot
\\ &\qquad\nonumber
\cdot\delta(\partial_0a_1-\partial_1a_0)\delta(\partial_2a_0-\partial_0a_2) Z[A_\mu + a_\mu]
.\end{align}
and expressing the constraints with Lagrange-multipliers gives
\begin{align}
    Z[A_\mu]
&= \nonumber
N_\Lambda N_1 N_2 \int Da_\mu DW_1 DW_2\ \det(\partial_0) Z[A_\mu+a_\mu]\cdot
    \\ &\nonumber\qquad
    \cdot\exp{\Big(-\frac{i}{2\pi}\int d^2xdt\ \varepsilon^{\mu\nu\lambda}W_j \partial_\nu a_\lambda\Big)}
\\ &=
N_\Lambda N_W \int Da_\mu DW_\mu\ \delta(W_0)\det(\partial_0) Z[a_\mu]\cdot
    \\ &\nonumber
    \qquad \cdot\exp{\Big(-\frac{i}{2\pi}\int d^2xdt\ \varepsilon^{\mu\nu\lambda}W_\mu \partial_\nu(a_\lambda-A_\lambda)\Big)}
\end{align}
The constraint on $W_0$ is a gauge fixing condition and corresponds to a Fadeev-Popov ghost \citep{Kapusta:2006th}
\begin{equation}
    \delta(W_0)\det(\partial_0) = \delta(\partial_\mu W_\mu)\det(\partial^2)
.\end{equation}
Furthermore, the effective action $(\ref{eq-surface-full})$ was obtained by expanding the fermionic partition function $Z[a_\mu]$ perturbatively in the field $a_\mu$, hence we get an additional prefactor corresponding to the free fermion
\begin{equation}
    Z[A_\mu] = Z_{\text{Fermi}}[0] \cdot \exp(iS_{\text{eff}}[A_\mu])
.\end{equation}
Hence, the bosonization formula with all prefactors included reads
\begin{multline}
   Z_{\text{Fermi}}[A_\mu]
   =
   Z_{\text{fermi}}[A_\mu=0]\cdot N_\Lambda N_W\int Da_\mu DW_\mu\ 
   \\  \delta(\partial_\mu W_\mu)\det(\partial^2)
   \exp\Big(iS_{\text{boson}}[a_\mu,W_\mu,A_\mu]\Big)
.\end{multline}
When setting $A_\mu=0$ everywhere,  the functional integral over the bosonic fields clearly has to be equal to unity. This is also apparent from the interpretation of $W_\mu$ as a Lagrange multiplier and $a_\mu$ as a gauge degree of freedom: integrating over $W_\mu$ constrains $a_\mu = \partial_\mu\Lambda$ and the Fadeev-Popov ghost cancels any spurious thermodynamics that might come from this integration. Then, $a_\mu$ is constrained to be a purely longitudinal field and the prefactor $N_\Lambda$ makes sure that integrating over it gives a finite result equal to unity.

In $1+1$ dimensions, the specific heat of the bosonic fields calculated in the manner of Equation $(\ref{eq-boson-energy})$ without keeping track of the ghost contributions will match the fermionic specific heat, but this may be a coincidence: The contributions from the Fadeev-Popov ghost have to cancel the bosonic parts and the fermionic determinant is generated when expanding the effective fermion action in perturbation theory.

\section{Refermionization}
\label{section-refermionization}
The usefulness of the bosonization procedure would be greatly increased if we could somehow express the original electron quantum field in terms of the collective bosonic excitations. Such a \emph{refermionization} program has been a great success in $1+1$ dimensions \citep{vonDelft:1998ff}, as it allows us to study strong electron interactions beyond perturbation theory. Two separate proposals for refermionization in $2+1$ dimensions have been put forth \citep{1979PhR....49..261L,Marino:1991ti}. We now show that these two approaches can be connected and we will demonstrate that a concrete expression for the fermion operator in terms of the field $W_\mu$ can indeed be constructed.

In the context of topological insulators, the possibility of refermionization was discussed in Ref.~\citep{Cho:2011bd}. We will very much follow this lead, although we have to make an important amendment concerning the physical interpretation of the bosonic fields first.

\def\q{\mathbf q}
\def\k{\mathbf k}
The first refermionization proposal by Luther \citep{1979PhR....49..261L} uses the so-called \emph{tomographic transformation} \citep{Sommerfield:1982we,1983PhRvD..28.2016A} to map $(2+1)$-dimensional bosons with a linear dispersion relation $E(\q) = v|\q|$ to  Dirac fermions. We will discuss this method in more detail in a moment, but our main concern right now is the following observation: It appears that we cannot identify these bosons with the charge or current density. Namely, if we look at the bosonized action for the fermions $(\ref{eq-surface-full})$ and use the identification $(\ref{eq-current})$ for the current $j^\mu$, we see that the correlation function for the current is essentially the Green's function $(\ref{eq-greens})$ of the square root $(-\square)^{1/2}$. But, bosons with a linear dispersion have an entirely different Green's function, namely, that for the wave operator $(-\square)$. This discrepancy is also apparent if we consider the excitation spectrum of the Dirac fermions, i.e.~the possible energies for creating an electron-hole pair. If an electron eigenstate with momentum $\k$ is excited to a state with momentum $\k+\q$, the energy difference is
\begin{equation}
    E(\q) = v|\k+\q| - v|\k|
.\end{equation}
Unlike in $1+1$ dimensions, the energy of this particle-hole excitation is not determined by the momentum $\q$ only; its possible values can be summarized in Figure~\ref{figure-spectrum}. The extremal parts where the electron momentum $\k$ is parallel to $\q$ do have a linear dispersion, but other 
contributions to the particle-hole continuum disperse in a different fashion and do not correspond to bosonic fields with a linear dispersion relation.

\begin{figure}
\begin{center}
\tikzsetnextfilename{figure-spectrum}
\begin{tikzpicture}[x=1.5cm,y=1.5cm]
    \fill[lightgray] (-1,1) -- (0,0) -- (1,1);
	\fill[lightgray] (-1,-1) -- (0,0) -- (1,-1);
    \draw[->] (-1.2,0) -> (1.2,0) node[right] {$|\q|$};
    \draw[->] (0,-1.2) -> (0,1.2) node[above] {$E(\q)$};
    \draw (-1,-1) -- (1,1);
    \draw (1,-1) -- (-1,1);
\end{tikzpicture}
\end{center}
\caption{The shaded area represents the possible energies $E(\q) = v|\k+\q| - v|\k|$ for particle-hole excitations of massless Dirac fermions. The hole has momentum $\k$.\label{figure-spectrum}}
\end{figure}
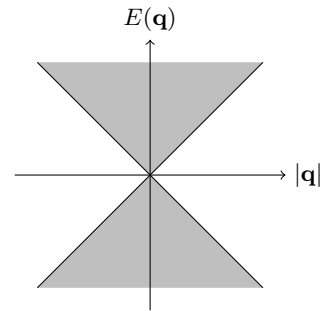

In other words, while Luther's construction yields an electron operator in terms of some bosonic fields, it appears that these bosonic fields cannot be related to physically relevant quantities in a straightforward manner because their correlation functions differ significantly. As a consequence, this also means that the refermionization of the topological insulator in Ref.~\citep{Cho:2011bd} suffers from the same problem of not allowing for a direct  relation of the bosonic fields to the charge density.

To make progress, we now consider the second refermionization proposal by Marino \citep{Marino:1991ti}. Essentially, this approach starts by considering the action $(\ref{eq-surface-full})$.  After integrating out the $a_\mu$ field, one   obtains an action expressed solely in the field $W_\mu$. Using the calculation from Appendix \ref{section-calculation}, in particular formula $(\ref{eq-surface-Euclidean})$, we obtain
\begin{multline}
    \label{eq-action-W}
    S[W_\mu] = \frac{1}{\pi^2(1+\alpha^2)}\int d^2xdt\
        \\ \left[-\frac12 W_{\mu\nu}(-\square)^{-1/2}W^{\mu\nu}
        - \alpha\varepsilon^{\mu\nu\lambda}W_\mu \partial_\nu W_\lambda\right]
.\end{multline}
We have written the transversal component using $W_{\mu\nu} = \partial_\mu W_\nu - \partial_\nu W_\mu$ to highlight the connection to the Maxwell term and abbreviated the prefactor of the Chern-Simons term with $\alpha=\frac{2\theta}{\pi^2}$.
The identification $(\ref{eq-current})$ of the current $j^\mu$ with the curl of the field $W^\mu$ suggests that the creation of an electron corresponds to the creation of a vortex in the vector field $W^\mu$. Hence, Marino proceeds to construct a vortex creation operator \citep{Marino:1991ti,1993AnPhy.224..225M} and identifies it with a fermion creation operator. However, these fermions need to be subjected to a transformation which is \emph{non-local in both time and space} in order to obtain the standard Dirac fermions.

We will not pursue this second path here. Rather, we wish to employ the tomographic transformation, but apply it to the physical field $W_\mu$ that encodes the charge density and current. The main idea taken from Marino's approach is that we have to apply a transformation that is non-local in both time and space. In particular, if we rescale the fields as
\def\W{\widetilde W}
\begin{equation}
    W_\mu := (-\square)^{1/4} \W_\mu
,\end{equation}
the action becomes
\begin{multline}
    \label{eq-action-rescaled-W}
    S[\W_\mu] = \int d^2xdt\ 
    \Big[-\frac12 \W_{\mu\nu}\W^{\mu\nu}
    \\ - \alpha\varepsilon^{\mu\nu\lambda}\W_\mu (-\square)^{1/2} \partial_\nu \W_\lambda\Big]
\end{multline}
up to an overall prefactor. It features a standard Maxwell term and some kind of ``non-local Chern-Simons term''. Both terms have the same order of derivatives, namely quadratic.

Now is a good time to apply the tomographic transformation. We will see that both the Maxwell and the Chern-Simons term, as well as their combination above transform very pleasantly. But first, let us review the mathematics of the tomographic representation, though our discussion will necessarily be brief. For additional details, we refer to Refs.~\citep{Sommerfield:1982we} and \citep{1983PhRvD..28.2016A}.

The basic idea of the tomographic transform is to think of a field $\phi$ in two-dimensional space not as a collection of amplitudes $\phi(\x)$ associated with  \emph{points} of space, but to think of them as amplitudes $\phi(\theta,y)$ associated with  \emph{lines} of the two-dimensional space. A line is specified by an angle $\theta\in[0,\pi)$ indicating direction and a number $y\in\R$ that indicates its signed distance to the origin, as illustrated in Figure~\ref{figure-tomographic}.
\def\n{\hat n}%
\begin{figure}
\begin{center}
\tikzsetnextfilename{figure-tomographic}
\begin{tikzpicture}[x=1.8cm,y=1.8cm]
    \def\y{0.7}
    \def\angle{25}
    \def\angleDistance{0.6}
    \draw[->] (-1,0) -> (1,0) node[right] {$x_1$};
    \draw[->] (0,-1) -> (0,1) node[above] {$x_2$};
    \draw [very thick,rotate=20] (\y,-1) -- (\y,1);
    \draw [gray,rotate=\angle] (-1,0) -- (1,0);
    \draw [gray,rotate=\angle,arrows=|<->|] (0,0) -- (\y,0)
        node[pos=0.5,above] {$y$};
    \draw [gray] (\angleDistance,0) arc (0:\angle:\angleDistance);
    \draw [gray] (0.5,0) node[above] {$\theta$};
\end{tikzpicture}
\end{center}
\caption{A line (solid) is represented by the angle $\theta$ of its normal vector and its distance from the origin $y$. The points $\x=(x_1,x_2)$ on this line are precisely those that fulfill the equation $y-\n\x = 0$.\label{figure-tomographic}}
\end{figure}
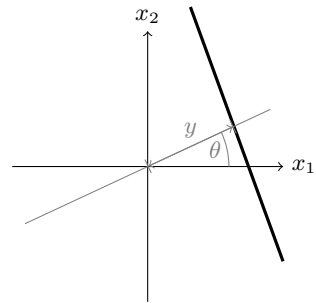
The amplitude associated with a line is essentially the sum of the amplitudes associated with the points on that line, i.e.~we define a unitary transformation
\begin{equation}
    (T\phi)(\theta,y) := \int d^2\x\ \delta^{1/2}(y-\n\cdot\x)\phi(\x)
\end{equation}
where $\n=\n(\theta)=(\cos \theta,\sin \theta)^T$ is a unit vector \emph{perpendicular} to the line and $\delta^{1/2}$ is a delta function
\begin{equation}
    \delta^{1/2}(y-\n\x)
        := \int_0^\infty dk\,k^{1/2} e^{-\eta k/2}(e^{ik(y-\n\x)}+e^{-ik(y-\n\x)})
\end{equation}
with the regulator $\eta\to  0$ in the end. The choice of $1/2$ for the exponent ensures that the transformation is unitary, so that for any two fields $\phi,\chi$, we have
\begin{equation}
    \int d^2x\ \phi(\x) \chi(\x) = \int d\theta dy\ (T\phi)(\theta,y)(T\chi)(\theta,y)
\end{equation}
where appropriate prefactors have been absorbed into the integration $\int d\theta dy$ over angles and distances. It was first noted by Radon \citep{Radon:1917tc} that this transformation can be inverted, so that the field $\phi(\x)$ can be reconstructed from its integrals over lines, though we will make no use of this fact here.

The most striking feature of the tomographic transform is how it maps derivatives of the fields. In particular, the Laplace operator becomes a simple square of the derivative $\partial_y$ independent of the angle $\theta$
\begin{align}
    T(\nabla \phi) &= \n \partial_y (T\phi)
\\    T(\Delta \phi) &= \partial_y^2 (T\phi)
.\end{align}
\def\W{\mathbf{W}}%
For vector fields like $\mathbf W = (W_1,W_2)^T$, it is convenient to define (spatially) longitudinal and transversal components as
\begin{align}
    W^L &:= \n\cdot T\W = n_1  TW_1 + n_2 TW_2
\\  W^T &:= \n\times T\W = n_2 TW_1 - n_2 TW_1
\end{align}
as this allows us to write (spatial) divergence and curl as
\begin{align}
    T(\nabla\cdot\W) &= \partial_y W^L
\\  T(\nabla\times \W) &= \partial_y W^T
.\end{align}
Together with unitarity, these transformation rules imply the following remarkable expressions for the Maxwell term
\begin{multline}
    \int d^2x\ \left[-\frac12 W_{\mu\nu}W^{\mu\nu}\right] =
        \int d\theta dy\ \\
        \Big[ (\partial_0W^T)^2 - (\partial_yW^T)^2 + (\partial_0W^L - \partial_y W_0)^2 \Big]
\end{multline}
and the Chern-Simons term
\begin{multline}
    \int d^2x\ \varepsilon^{\mu\nu\lambda}W_\mu \partial_\nu W_\lambda = \int d\theta dy\ 2W^T(\partial_0W^L - \partial_y W_0)
.\end{multline}
In particular, the combinations $W^T$ and $W_{0L} := (\partial_0W^L - \partial_y W_0)$ are invariant under gauge transformations $W_\mu \to W_\mu + \partial_\mu \Lambda$. To keep notation simple, we have used the same symbol $W$ for both the original field and its tomographic transform here.


\def\W{\widetilde{W}}%
The action $\eqref{eq-action-rescaled-W}$ for the field $\W_\mu$ can now be written in tomographic space as
\begin{multline}
    S[\W_\mu] = \int d\theta dydt\ 
        \Big[(\partial_0 \W^T)^2 - (\partial_y \W^T)^2 + (\W_{0L})^2
            \\ - 2\alpha\W^T (\partial_y^2 - \partial_0^2)^{1/2} \W_{0L} \Big]
.\end{multline}
Shifting
\begin{equation}
    \widehat W_{0L} = \W_{0L} + \alpha (\partial_y^2 - \partial_0^2)^{1/2}\W^T
\end{equation}
\def\W{\widehat{W}}%
absorbs the Chern-Simons term in the component $\W_{0L}$ whose dynamics are trivial. We obtain
\begin{multline}
    S[\W^T,\W_{0L}] = \int ddt\theta dy\ \Big[ 
        (\W_{0L})^2 
        \\ + (1-\alpha^2)\W^T(\partial_y^2 - \partial_0^2) \W^T
        \Big]
\end{multline}
Discarding or integrating out the uninteresting $\W_{0L}$ component, we see that the action for the $\W^T$ field is precisely the action needed for refermionization in $1+1$ dimensions. The equation of motion is simply the wave equation
\begin{equation}
    (\partial_y^2 - \partial_0^2) \W^T = 0
\end{equation}
and we can decompose the field into left- and right-moving parts
\begin{equation}
    \W^T(\theta,y,t) = W_{\text{left}}(\theta,y+t) + W_{\text{right}}(\theta,y-t)
.\end{equation}
Hence, for every angle $\theta$, we can apply the refermionization procedure familiar from $(1+1)$-dimensional bosonization \citep{vonDelft:1998ff} to obtain fermion fields as a normal ordered version of
\begin{equation}
    \label{eq-fermion-field}
    \tilde\psi_{\pm}(\theta,y) = C \exp\left(i\sqrt{\pi}W_{\text{left/right}}(\theta,y,t)\right) \hat O_\theta
\end{equation}
where $C$ is some constant and $\hat O_\theta$ refers to Klein factors that are needed to obtain the correct anticommutation relations. Undoing the tomographic transform yields a massless field of $(2+1)$-dimensional Dirac fermions again, though we refer to Refs.~\citep{1983PhRvD..28.2016A} and \citep{1979PhR....49..261L} for the details on this construction.

To summarize, the tomographic representation allows us construct a fermion operator $\psi$ from the field $W_\mu$ with the same Green function as the $(2+1)$-dimensional massless Dirac fermion. From this, we can see that the current $\bar \psi\gamma^\mu \psi$ of this new fermion has the same correlation functions as the current $j^\mu = \varepsilon^{\mu\nu\lambda}\partial_\nu W_\lambda$ of the original fermion field. Hence, we propose to identify the original fermion field with the newly constructed fermion.

As we have already indicated, one way in which this refermionization procedure may be useful is the study of electron interaction. For, instance a Hubbard-like local interaction of the form $g j^0(x) j^0(x)$ can be incorporated into the bosonic action \eqref{eq-action-W} via an additional term
\begin{equation}
    S_{\text{int}}[W] = \int d^2xdt\,\Big[-g (\partial_1W_2 - \partial_2W_1)^2\Big]
.\end{equation}
The main point is that this term is still quadratic in the field $W_µ$, which means that the correlation functions $\<W_µ(x) W_ν(x')\>$ can be calculated exactly with a Gaussian integration. By virtue of the refermionization formula \eqref{eq-fermion-field}, this in turn can be used to calculate the Greens function $\<\tilde\psi^\dagger(x) \tilde\psi(x')\>$ in the presence of interaction. Of course, we have to keep in mind that, unlike in $1+1$ dimensions, the bosonic action \eqref{eq-action-W} is only an approximation, so this will not be an exact solution of the interacting electron problem.

\section{Summary and Discussion}
\label{section-summary}
Guided by the principle of anomaly cancellation between bulk and surface, we have constructed a topological field theory for the TRS topological insulator in three space dimensions. We have incorporated the electron dynamics into the bosonic surface theory by considering the effective electromagnetic response obtained from the fermionic Dirac Hamiltonian. This bosonic theory does not reproduce the fermionic thermodynamics quantitatively, but we have shown that this discrepancy is accounted for by normalization factors of auxiliary functional integrals. Using the tomographic transform, we have proposed a representation of the fermionic field in terms of the bosonic field, although the non-locality of the bosonic action forces this representation to be highly non-local as well.

Our approach is very much inspired by the construction of a topological field theory for the TRS topological insulator in three dimensions in Ref.~\citep{Cho:2011bd}. In contrast to Ref.~\citep{Cho:2011bd}, however, we have argued that the TRS insulator is described, perhaps counterintuitively, by a surface theory that explicitly breaks time-reversal symmetry, even though the combination of bulk and surface is again time-reversal invariant. Also, we have found that the effective dynamics of the charge density on the surface are highly non-local in both space and time, as can be seen by examining the spectrum of particle-hole excitations.

Similar to the construction of topological field theories in Ref.~\citep{2013PhRvB..87h5132C}, the method of functional bosonization has allowed us to  derive systematically a field theory from the response to an external field. However, our main guiding principle is not just the electromagnetic response, but the cancellation of quantum anomalies, which has allowed us to predict the existence of surface states from the topological bulk theory. Moreover, we were able to identify the appropriate degrees of freedom needed for a topological field theory of the quantum spin Hall effect by discussing the axial anomaly in Appendix \ref{section-axial-anomaly}.

We believe that the parity anomaly may have additional significance in the context of refermionization. The Chern-Simons term is known to induce a statistical transmutation \citep{1988MPLA....3..325P,1992IJMPB...6...25Z}, mapping bosons to fermions and vice-versa. Hence, the presence of a Chern-Simons term in the effective bosonic action $\eqref{eq-surface-full}$ is a strong indication that the bosonic surface theory actually describes fermionic excitations. The Klein factors in the refermionization proposal $\eqref{eq-fermion-field}$ are yet to be constructed explicitly in terms of the bosonic field $W_\mu$, but they may be related to this statistical transmutation.

Acknowledgements: We would like to thank Ady Stern for collaborating with us in the initial stage of this project, Daniel Scherer and Mirco Milletari for helpful discussions, Norman Metzner for feedback on the manuscript, the anonymous reviewers for valuable comments and the BMBF for financial support.

\appendix
\section{Field Integral Calculations with the Chern-Simons Term}
\label{section-calculation}
We now introduce some tools which facilitate the calculation of  functional integrals involving the $(2+1)$-dimensional Chern-Simons term. The key convenience is that there exist \citep{Barci:1996bw} two orthogonal projection matrices $P^+,P^-$ that project a vector $a_\mu$ onto its (spacetime) transversal components
\begin{equation}
    (P^+ + P^-)_{\mu\nu} a_\nu
        = \left(\delta_{\mu\nu} - \frac{q_\mu q_\nu}{q^2}\right) a_\nu
        = a^T_\mu
\end{equation}
but are also related to the Chern-Simons
 \footnote{Recall that Minkowski spacetime and Euclidean spacetime are related by a Wick rotation $it \to \tau$ and the Minkowski respectively Euclidean actions are related by $iS_M = -S_E$. Hence, the Chern-Simons action in Minkowski spacetime $S_M[a_\mu] = \int d^2xdt\ a_\mu\varepsilon^{\mu\nu\lambda}\partial_\nu a_\lambda$ becomes $S_E[a_\mu] = -i\int d^2xd\tau\ a_\mu\varepsilon_{\mu\nu\lambda}\partial_\nu a_\lambda$ in Euclidean spacetime, picking up a prefactor of $(-i)$.}
term via
\begin{equation}
    a_\mu |q| (iP^+ - iP^-)_{\mu\nu} a_\nu = \varepsilon_{\mu\alpha\nu} a_\mu q_\alpha a_\nu
.\end{equation}
These two matrices are readily obtained from the ``Chern-Simons matrix''
\begin{equation}
    C_{\mu\nu} = -i\varepsilon_{\mu\alpha\nu}\frac{q_\alpha}{|q|}
\end{equation}
as the relations
\begin{align}
    C^2 &= C_{\mu\beta}C_{\beta\nu} = \delta_{\mu\nu} - \frac{q_\mu q_\nu}{q^2}
\\
    C^3 &= C_{\mu\beta}C_{\beta\gamma}C_{\gamma\nu} = C_{\mu\nu} = C
\end{align}
allow us to write
\begin{multline}
    P^{\pm}_{\mu\nu} = \frac12 (C^2_{\mu\nu} \pm C_{\mu\nu})
    , \quad (P^{\pm})^2 = P^{\pm}
    , \quad P^{+} P^{-} = 0
.\end{multline}

With these tools, we can write the surface action $(\ref{eq-surface-full})$ in Euclidean momentum space as
\begin{multline}
    \label{eq-surface-Euclidean}
    S[a_\mu,W_\mu] = \sum_{q} \Big[
        -\frac1{2\pi} a_\mu(-q)|q|(iP^+ - iP^-)_{\mu\nu} W_\nu(q)
        \\ +\frac{1}{16} a_\mu(-q)|q|\big((1+i\alpha)P^+ + (1-i\alpha)P^-\big)_{\mu\nu} a_\nu(q)
    \Big]
\end{multline}
using the notation $\alpha=\frac{2\theta}{\pi^2}$.

For instance, it is now straightforward to integrate out the field $a_\mu$ and obtain the action expressed in terms of the field $W_\mu$ alone. Namely, the linear shift
\begin{equation}
    a_\mu \to a_\mu + \frac{4}{\pi}\left(\frac{i}{1+i\alpha}P^+ - \frac{i}{1-i\alpha}P^- \right)_{\mu\nu} W_\nu
\end{equation}
will complete the square and leave the residual action
\begin{multline}
    S[W_\mu] = \frac1{\pi^2(1+\alpha^2)}\sum_q 
        \\ W_\mu |q|\left(P^+ + P^- - i\alpha(P^+ - P^-) \right)_{\mu\nu} W_\nu
\end{multline}
in Euclidean momentum space.

\section{Axial Anomaly and the Quantum Spin Hall Insulator}
\label{section-axial-anomaly}
We now want to come back to the discussion of quantum anomalies from Section~\ref{section-quantum-anomalies} and explain the edge states of the TRS topological insulator in two space dimensions in terms of the \emph{axial anomaly} \citep{Jackiw:2008fg,Fujikawa:2012uc}. For concreteness, we will focus on the special case of the quantum spin Hall insulator \citep{Kane:2005hl}, and assume that the electrons in the planar bulk carry a conserved quantum number, namely the spin in perpendicular direction,  $S_z$. This model system is representative for TRS topological insulators \citep{Kitaev:2009vc}, but we  admit that it is not entirely clear how to find a substitute for the spin current in the general case. Still, we wish to argue that it beautifully captures the topological essence of the situation.

The quantum spin Hall insulator can  be understood as two opposing copies of the quantum Hall effect: The spin-up electrons experience a Hall effect with conductivity $\sigma_{xy}$, while the spin-down electrons are subject to a Hall conductivity $-\sigma_{xy}$, i.e.~they move in the opposite direction. Time reversal corresponds to interchanging the spin up and spin down directions, and the system is therefore invariant under time-reversal. Since the electrons move in opposite directions, the total electric Hall current vanishes, but we can observe a spin Hall current.

Again, the key feature of the quantum spin Hall insulator is that the edge of a finite sample features gapless electron states. Since spin is conserved, we could use our knowledge of the quantum Hall effect and the chiral anomaly to predict that the edge has two counterpropagating modes: the spin-up electrons moving in one and the spin-down electrons moving in the other direction along the edge. However, anticipating a generic topological insulator where spin is not conserved, we want to answer the following question: How can we obtain the \emph{combined edge state} from a quantum anomaly without separating the electron into independent spin components?

The main observation is that the combination of the two edge modes is equivalent to the $(1+1)$-dimensional Dirac fermion,
which is known to suffer from the \emph{axial anomaly} \citep{Jackiw:2008fg,Fujikawa:2012uc}. Essentially, the axial anomaly is a competition between the charge current $j^\mu = \bar \psi\gamma^\mu \psi$ and the spin (or \emph{axial}) current $j^\mu_B = \bar \psi\gamma^\mu\gamma_5 \psi$: while the classical action for the Dirac fermion suggests that both currents satisfy the continuity equation, the quantum theory actually predicts that the currents cannot be  \emph{conserved simultaneously}, we can choose between either $\partial_\mu\<j^\mu\> = 0$ or $\partial_\mu \<j^\mu_B\> = 0$, but we can never have both. The underlying reason is that the quantum theory needs to be regularized and the inevitable choice corresponds to a choice of regularization.
We will explain this in more detail in Appendix \ref{section-quantum-regularization}, for now let us explore the consequences. In particular, the necessity to choose also makes it clear that we cannot predict the edge modes from considering the charge current or the spin current alone, because each of them can  be conserved individually and we would not observe a defect.

From this point of view, we must consider an effective bulk action $S_{\text{eff}}[A_\mu,B_\mu]$ that depends on both the electromagnetic gauge field $A_\mu$ and a (fictitious) axial gauge field $B_\mu$ so that we can probe both the charge current via $j^\mu = \frac{\delta S[A,B]}{\delta A_\mu}$ and the axial (spin) current via $j^\mu_B = \frac{\delta S[A,B]}{\delta B_\mu}$. Essentially, we think of it as a generating functional for the two currents. Invariance of the action under gauge transformations $A_\mu \to A_\mu + \partial_\mu\Lambda$ or $B_\mu \to B_\mu + \partial_\mu \Theta$ means that the charge current, respectively the axial current, is conserved. The effective action is easily engineered from the Chern-Simons theory of the quantum Hall effect \citep{Wen:1992cf}. Representing the spin-up electrons by a collective field $a$ and the spin-down electrons by a collective field $b$, we couple the electromagnetic field with the same sign, but the axial field with opposite signs
\begin{multline}
    S[a,b,A_\mu,B_\mu] = \frac{1}{2\pi} \int d^2x dt \Big[
    \\ -\frac1{4\pi \sigma_{xy}} \varepsilon^{\mu\nu\lambda} a_\mu\partial_\nu a_\lambda + \varepsilon^{\mu\nu\lambda} A_\mu\partial_\nu a_\lambda + \varepsilon^{\mu\nu\lambda} B_\mu\partial_\nu a_\lambda
    \\ +\frac1{4\pi \sigma_{xy}} \varepsilon^{\mu\nu\lambda} b_\mu\partial_\nu  b_\lambda + \varepsilon^{\mu\nu\lambda} A_\mu\partial_\nu b_\lambda - \varepsilon^{\mu\nu\lambda} B_\mu\partial_\nu b_\lambda \Big]
.\end{multline}
Integrating out the matter fields, or simply noting that the spin-up electrons basically see a total gauge potential of $A_\mu+B_\mu$ whereas the spin-down electrons see a total gauge potential of $-A_\mu + B_\mu$, we obtain the effective action
\begin{equation}
    \label{eq-axial-bulk-AB}
    S_{\text{eff}}[A_\mu,B_\mu] = \sigma_{xy}\int d^2xdt\ \Big[ \varepsilon^{\mu\nu\lambda} A_\mu\partial_\nu B_\lambda + \varepsilon^{\mu\nu\lambda} B_\mu\partial_\nu A_\lambda \Big]
.\end{equation}

As in our previous discussions, for periodic boundary conditions, the effective action is invariant under gauge transformations of the fields $A_\mu$ and $B_\mu$. In fact, integrating partially, we can write the action in many different ways
\begin{align}
    S_{\text{eff}}[A_\mu,B_\mu]
&= \label{eq-axial-bulk-A}
2\sigma_{xy}\int d^2xdt\ \varepsilon^{\mu\nu\lambda} B_\mu\partial_\nu A_\lambda
\\ &= \label{eq-axial-bulk-B}
2\sigma_{xy}\int d^2xdt\ \varepsilon^{\mu\nu\lambda} A_\mu\partial_\nu B_\lambda
\end{align}
so that the action is manifestly gauge invariant in one field, $A_\mu$ in the first equation, because the action only depends on its curl, and gauge invariant in the other field, $B_\mu$ in the first equation, because of the periodic boundary conditions. In the second equation, the manifestation of the gauge invariance is reversed.

Now, if we try to restrict the bulk action to a bounded domain, say to the region $x_1>0$, we can choose between many different forms like $(\ref{eq-axial-bulk-AB})$,$(\ref{eq-axial-bulk-A})$ and $(\ref{eq-axial-bulk-B})$ for the bulk action. While puzzling at first, this choice is precisely the ambiguity that the axial anomaly allows: we may choose to conserve either the charge or the spin current! When choosing the form $(\ref{eq-axial-bulk-A})$ where gauge invariance of the electromagnetic field is manifest, the charge current will be conserved, but the spin current will be anomalous and must be canceled by a corresponding edge mode
\begin{align}
    \partial_\mu j^\mu_{\text{bulk}} &= 0
\\    \partial_\mu j^\mu_{B,\text{bulk}} &= 2\sigma_{xy} \delta(x_1) \varepsilon^{1\nu\lambda} \partial_\nu A_\lambda
\label{eq-spin-current-anomaly}
.\end{align}
Similarly, the form $(\ref{eq-axial-bulk-B})$ must have an edge mode that cancels an anomalous charge current while the spin current is conserved. Mixed forms are also possible.

To summarize, we have shown that the bulk action(s) $(\ref{eq-axial-bulk-AB})$ predicts the edge states of the quantum spin Hall insulator via the axial anomaly. In particular, it correctly reproduces the ambiguity inherent in the axial anomaly and shows that one must consider both charge and spin current in combination to predict the edge state. Again, it is straightforward to construct a corresponding topological field theory via bosonization as presented in Section~\ref{section-topological-field-theory}.

With regards to the ambiguity of the axial anomaly, in a condensed matter physics setting, it seems likely that the charge current is conserved at the edge and that the spin current must carry the defect. However, as we discuss in Appendix~\ref{section-quantum-regularization}, the response of the electronic edge state is not well-defined, it only becomes physically meaningful in conjunction with the bulk response. But the total system is free of ambiguities as the anomalies cancel.
For the quantum spin Hall insulator with a conserved spin component, the axial current is the physical spin current. For the more general case without spin conservation, the axial anomaly was argued to be observable in the form of pair switching \cite{Ringel:2012uo}.

\section{A Quantum Anomaly Primer}
\label{section-quantum-regularization}
To make our discussion self-contained, we now present a very short general introduction to the subject of quantum anomalies in quantum field theory. For a more thorough introduction, we refer to Refs.~\citep{Jackiw:2008fg} and \citep{Fujikawa:2012uc}. In particular, we want to highlight the origin of the parity anomaly for the Dirac fermion in $2+1$ dimensions, which plays a key role in our derivation of the topological field theory in Section~\ref{section-quantum-anomalies}. It is also important for the question as to whether the surface states feature an experimentally measurable quantum Hall effect or not that we answered in Section \ref{section-surface-dynamics}. Likewise, we want to explain the ambiguity inherent in the edge states of the quantum spin Hall system discussed in Appendix~\ref{section-axial-anomaly}.

In quantum field theory, a quantum anomaly is the phenomenon that a symmetry may be present in the classical Lagrangian, but is lost once the theory is quantized. The underlying reason for this defect is that the quantum theory needs to be regularized and it is impossible to find a regularization scheme that could preserve the symmetry. This impossibility is usually related to topological obstructions. We now want to explain the issue of regularization in more detail.

For concreteness, consider again the partition function \eqref{eq-partition-dirac} of the massless Dirac fermion in $2+1$ spacetime dimensions coupled to an electromagnetic gauge potential $A_\mu$, expressed in Euclidean spacetime
\begin{equation}
    Z[A_\mu] = \int D\bar \psi D\psi \exp\left(-\int d^2xd\tau\ \bar \psi \D \psi \right)
.\end{equation}
We have simplified the notation by introducing the \textit{Dirac operator} $\D = \D[A_\mu] := \gamma^\mu(\partial_\mu - iA_\mu)$. In Euclidean spacetime, the Dirac operator is hermitian, $\D^\dagger = \D$ and we can expand the electron field in terms of an eigenbasis
\begin{align}
    \nonumber
    \D \phi_n(x) &= \lambda_n \phi_n(x)
\\    \psi(x) &= \sum_n a_n \phi_n(x), \quad \bar\psi(x) = \sum_n \bar a_n \phi_n(x)
\end{align}
where $a_n,\bar a_n$ are Grassmann-valued coefficients. Note that the Dirac operator acts on fields depending in spacetime coordinates, so these are not the eigenstates of the corresponding Hamiltonian (though they can be related). Expressing the field integral in terms of this eigenbasis, we obtain
\begin{align}
    \label{eq-product-eigenvalues}
    Z[A_\mu]
    \nonumber
    &= \prod_n d\bar a_n da_n  \exp\left(-\sum_n \lambda_n \bar a_n a_n\right)
    \\
    &= \prod_n \lambda_n = \det \D
\end{align}
which is the determinant of the Dirac operator as expected.

Unfortunately, this expression for the determinant is not well-defined, as the eigenvalues $\lambda_n$ of the Dirac operator grow without bounds. We have to \emph{regularize} them in some way to obtain a well-defined quantum theory of the Dirac fermion in $2+1$ dimensions. There are several ways to regularize the infinite product of eigenvalues. For instance, we can cut off the product and only consider the first $N$ eigenvalues
\begin{equation}
    Z_{\text{cut-off}}[A_\mu] := \prod_{n=1}^N \lambda_n
.\end{equation}
Another option more suitable for perturbative calculations is Pauli-Villars regularization
\begin{equation}
    Z_{\text{Pauli-Villars}}[A_\mu] := \lim_{M\to\infty} \frac{\det \D}{\det(\D + M)}
.\end{equation}
The division by a determinant corresponds to the introduction of a bosonic field that helps to regularize the theory but does not contribute to the dynamics as its mass gap $M$ is sent to infinity.

Now, the classical action of the $(2+1)$-dimensional Dirac fermion was invariant under both gauge transformations and time-reversal symmetry, but it turns out that any regularization of the path integral \eqref{eq-product-eigenvalues} must necessarily break one of these symmetries. This was shown in Refs.~\citep{Redlich:1984ck} and \citep{Redlich:1984hu} using a topological argument with homotopy groups. (In sense, this breaking of symmetry can be attributed to an asymmetry of the path integral measure $\prod_n da_n d\bar a_n$.) Note that the two regularization schemes mentioned above preserve the gauge symmetry because they only depend on the eigenvalues $\lambda_n$ of the Dirac operator, which are invariant under gauge transformations. In contrast, performing a na\"ive regularization by introducing a cut-off in momentum space would break gauge invariance, as a gauge transformation $A_\mu \to A_\mu + \partial_\mu \Lambda$ can introduce arbitrary high-momentum variations into the fields.

It may seem unusual to consider several regularization schemes, because in a condensed matter setting, the physically correct regularization scheme for the fermion determinant is lattice regularization which we expect to preserve both symmetries. We will discuss this in a moment, but the gist of the problem is that it is actually not possible to give a $(2+1)$-dimensional lattice model for a single Dirac fermion due to the Nielsen-Ninomiya theorem \citep{1981NuPhB.185...20N}, so this regularization scheme is unavailable.

If we adopt Pauli-Villars regularization, the effective action for the $(2+1)$-dimensional Dirac fermion will acquire an anomalous term that breaks time-reversal symmetry. Let us consider the slightly more general situation of a massive Dirac fermion with mass $m$. The discussion about regularization applies essentially unchanged and the partition function is
\begin{align}
    \nonumber
    Z_{\text{PV}}[m,A_\mu]
    &= \int D\bar \psi D\psi \exp\left(-\int d^2xd\tau\, (\bar \psi \D \psi + m \bar \psi \psi) \right)
    \\
    &:=
    \lim_{M\to\infty} \frac{\det(\D+m)}{\det(\D + M)}
.\end{align}
Taking logarithms and expanding the determinant to second order in the field $A_\mu$, we obtain the effective action
\begin{align}
    S_{\text{eff}}[m,A_\mu]
    &=
    -\ln Z_{\text{PV}}[m,A_\mu]
\nonumber    \\ &=
    -\lim_{M\to\infty} \left( \ln \det(\D+m) - \ln\det(\D + M) \right)
\nonumber    \\ &=
    \lim_{M\to\infty} \left(\frac12 A_\mu \Pi^{\mu\nu}(m,M) A_\nu \right) + \mathcal{O}(A^3_\mu)
\end{align}
where the polarization tensor is given by
\begin{align}
    \label{eq-polarization-integral}
    \Pi^{\mu\nu}(m,M)
    &= \Pi^{\mu\nu}(m) - \Pi^{\mu\nu}(M)
    \\
    \nonumber
    &= \Tr[S(m) \gamma^\mu S(m)\gamma^\nu - S(M) \gamma^\mu S(M) \gamma^\nu] 
\end{align}
and the Greens function of the free electron is
\begin{equation}
    S(m) = (\D+m)^{-1} = (\gamma^\mu\partial_\mu + m)^{-1}
.\end{equation}
In the formula for the polarization tensor, the trace corresponds to an integral over spin and spacetime momentum. It would be divergent in the ultraviolet if it were not for the regularization that we subtract a contribution with large mass $M$. This cures the divergence, but the price is that it also introduces a spurious term that looks like a quantum Hall effect -- the parity anomaly.

Performing the trace in Equation \eqref{eq-polarization-integral} gives the effective action \eqref{eq-surface-massive} for the massive Dirac fermion in $2+1$ dimensions. It includes both the parity anomaly and an additional quantum Hall effect related to the mass $m$. These two terms look very similar because they both arise from a mass, but keep in mind that the parity anomaly is solely an artifact of the regularization. A different choice of regularization, for instance one that breaks gauge invariance, would introduce a different term. However, the main point we make in this paper is that the form of the anomalous term does not really matter: it is canceled by the bulk anyway because the total system must preserve all symmetries.

A similar discussion applies to the axial anomaly and the edge state of the quantum spin Hall effect that we discussed in Appendix~\ref{section-axial-anomaly}. The Dirac operator in $1+1$ dimensions now includes both the electromagnetic $A_\mu$ and the axial gauge potential $B_\mu$, but its determinant has to be regularized as before. Again, a topological argument \citep{Fujikawa:2012uc} shows that one of the gauge symmetries must be broken, and the choice of regularization determines which one. For instance, a regularization that preserves the charge current yields the partition function of the edge states
\begin{multline}
    \ln Z_{A}[A_\mu,B_\mu] = i\sigma_{xy} \int dxdt\,
        \\ 
        \Big( -\frac12 A^T_\mu A^{T,\mu} - \frac12 B^T_\mu B^{T,\mu}
        + 2\epsilon^{\mu\nu}B^L_\mu A^T_\nu \Big)
\end{multline}
which corresponds to Equation~\eqref{eq-spin-current-anomaly}. Here, $A^L_\mu= \partial_\mu \partial^\nu (\square)^{-1} A_\nu$ is the longitudinal part and $A^T_\mu = A_\mu - A^L_\mu$ is the transversal part of the gauge field $A_\mu$; similarly for the field $B_\mu$. In contrast, a regularization that preserves the spin current yields
\begin{multline}
    \ln Z_{A}[A_\mu,B_\mu] = i\sigma_{xy} \int dxdt\,
        \\ 
        \Big( -\frac12 A^T_\mu A^{T,\mu} - \frac12 B^T_\mu B^{T,\mu}
        - 2\epsilon^{\mu\nu}A^L_\mu B^T_\nu \Big)
.\end{multline}
Mixed forms are also possible.

All these seem to be valid results for the partition function of the edge state of the quantum spin Hall isolator. However, in a condensed matter setting, the physically correct choice of regularization is lattice regularization, and we would expect that the ambiguity is resolved in favor of one physical choice of partition function. But, the problem is as follows: There is no one-dimensional lattice theory that reproduces the $(1+1)$-dimensional Dirac fermion, due to the Nielsen-Ninomiya fermion doubling theorem \citep{1981NuPhB.185...20N,Fujikawa:2012uc}. That does not mean that lattice regularization is impossible, evidently, we can realize the $(1+1)$-dimensional Dirac fermion as the low-energy theory at the edge of a quantum spin Hall system on a lattice. However, in a sense, additional contributions of the bulk lattice are now responsible for regularizing the theory.

Another way of interpreting the difficulties with lattice regularization is that only theories \emph{without} quantum anomalies admit a lattice model. If the continuum theory has a quantum anomaly, then a lattice model must include bulk or other contributions that cancel the anomaly. This is another viewpoint on our main message: Effective actions derived from continuum models for surface and bulk may include quantum anomalies, which may even be ambiguous, but the total system has a lattice model and the anomalous terms must cancel.

\bibliography{bosonization}

\end{document}